\documentclass[
aps,
 ,twocolumn
]{revtex4}
\usepackage{graphicx}

\begin{document}

\title{Quantum Error Correction of Continuous Variable States against Gaussian Noise}

%
%
%
\author{T.C.Ralph}\affiliation{Centre for Quantum Computation and Communication Technology, \\
School of Mathematics and Physics, University
of Queensland, St Lucia, Queensland 4072, Australia}
%
%
%


\date{\today}

\begin{abstract}

We describe a continuous variable error correction protocol that can correct the Gaussian noise induced by linear loss on Gaussian states. The protocol can be implemented using linear optics and photon counting. We explore the theoretical bounds of the protocol as well as the expected performance given current knowledge and technology.

\end{abstract}


\maketitle


\vspace{10 mm}

{\it Introduction}: Continuous variable quantum information protocols use quantum operations and measurements acting on states with continuous eigenvalue spectra to perform quantum information tasks such as quantum teleportation, quantum key distribution and quantum processing \cite{LOO05}. An attraction of continuous variable protocols is that many require only Gaussian states, operations and measurements \cite{note1} - all of which can be implemented deterministically in optics with current technology. However, many applications also require the ability to error correct the quantum states in order to realize their full potential. 

Recently it has been proven that error correction of Gaussian noise, imposed on Gaussian states, using Gaussian operations is impossible \cite{Nis09}. This is a significant result as Gaussian noise is the most common source of errors for continuous variable states. It is thus of considerable interest to determine whether additional, non-Gaussian resources, can be employed to allow error correction of continuous variable states against Gaussian noise. Here we answer this question in the affirmative by describing an error correction protocol that is effective against the Gaussian noise produced by loss. Although in principle our protocol could be applied to any physical architecture, we will focus particularly on optics given its experimental relevance. The only additional non-Gaussian operation required for our protocol is photon counting.

A Gaussian continuous variable error correction protocols based on a direct generalization of the Shor 9-qubit error correction code \cite{Sho95} has been developed \cite{Llo98}, \cite{Bra98b} and demonstrated experimentally \cite{Aok09}. This code can correct a large range of non-Gaussian errors but not Gaussian errors. Other protocols for correcting more specific types of non-Gaussian noise imposed on Gaussian states, using Gaussian operations have also been proposed and demonstrated \cite{Nis08,Las10}. Methods for correcting Gaussian noise imposed on specific non-Gaussian code states have also been described \cite{Got01,Lun08}.

We consider a situation in which we wish to transmit an ensemble of quantum states through a channel of loss $\eta$ (see Fig.1(a)). The loss inevitably 
couples the system to the environment and reduces the distinguishability of the states, producing errors in any quantum information encoded in the ensemble. Successful error correction should reduce the effective loss on the channel thus leading to a lower error rate .
Instead of considering error correction based on error correction codes such as those discussed above, we will consider error correction based on the distillation of entanglement \cite{Ben96}, and the subsequent use of the distilled entanglement for teleportation (see Fig.1(b)). Distillation of continuous variable entanglement is known to be possible using photon counting \cite{Bro03, LUN09}. Here we will use a convenient distillation approach based on heralded noiseless linear amplification that has been demonstrated recently \cite{Xia10}.
\begin{figure}[htb]
\begin{center}
\includegraphics*[width=8cm]{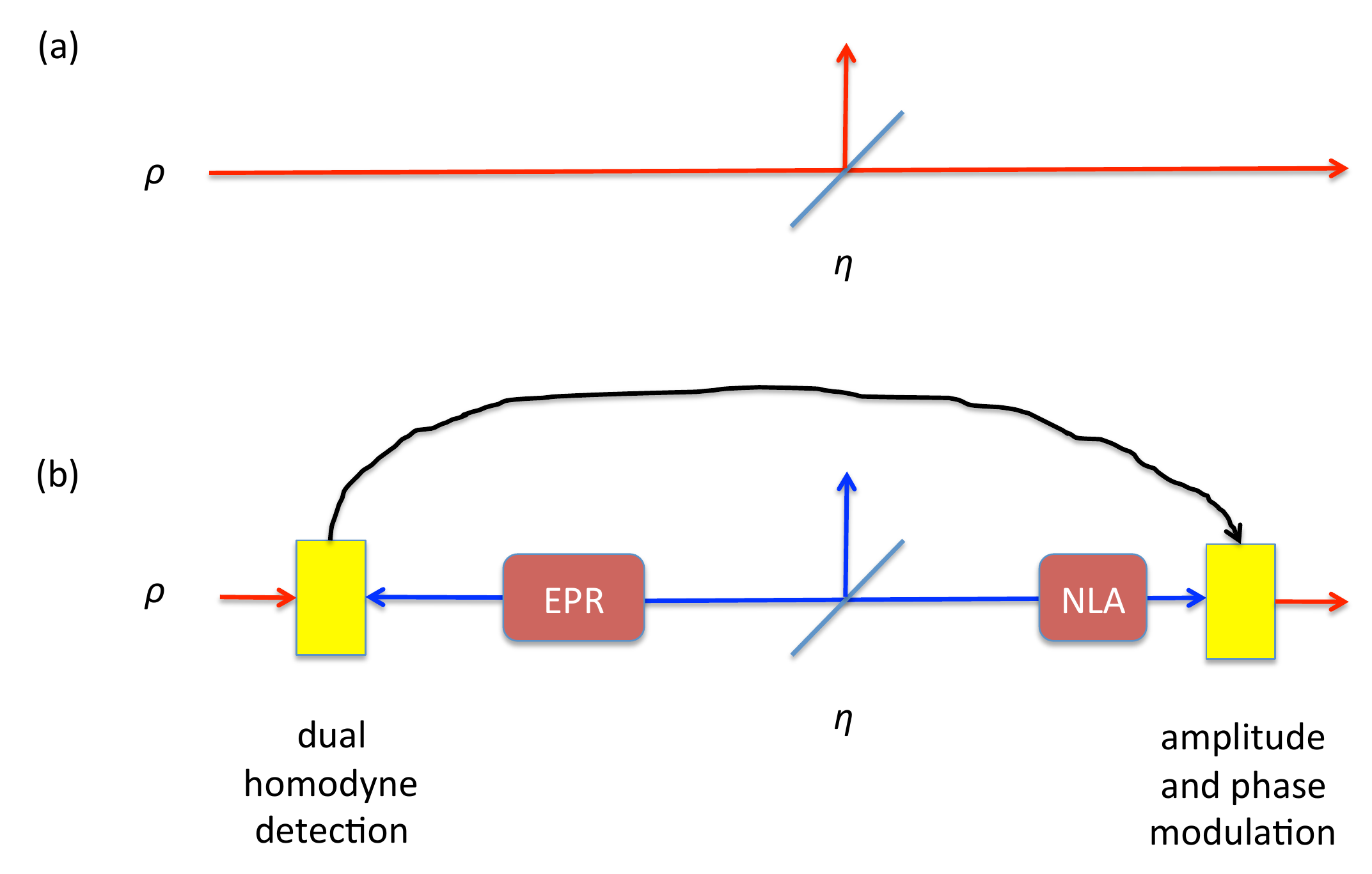}
\caption{Example of a protocol for error correcting Gaussian states against Gaussian noise.  (a) An ensemble of states $\rho$ is being transmitted through a channel of loss $\eta$. The loss inevitably causes errors to the quantum information encoded in the ensemble through the coupling to the environment. (b) Now continuous variable entanglement (EPR) is distributed through the lossy channel and post amplification of gain $G$  is attempted via a noiseless amplifier (NLA). Noiseless amplification is non-deterministic so most attempts fail. These are discarded. When an attempt succeeds the entanglement is used to teleport the input to the output by making dual homodyne measurements at the input and sending the results to the output where they are used to displace the output state. If the gain of the NLA is sufficiently large and the gain of the output displacement is chosen correctly, the effective channel is a lossy channel of higher transmission than the direct channel, and hence the output state contains less errors.}
\label{fig1}
\end{center}
\end{figure}

Although a well known equivalence between error correction and distillation exists for discrete variables the situation is not so straightforward for continuous variables. The problem is that continuous variable entanglement is only strictly maximal in the limit of infinite energy. The effect of non-maximal entanglement is to add noise in the teleportation protocol, and thus potentially compromise any error correction achieved via the distillation. However, here we show that by modifying the teleportation protocol, error correction can be achieved in the absence of maximal entanglement. 

{\it Teleportation}: Continuous variable teleportation utilizes shared Einstein, Podolsky, Rosen (EPR) entanglement (also known as two-mode squeezing) and homodyne detection to transmit quantum field states between two locations \cite{BRA98}. One arm of the entanglement and the state to be transmitted is held by Alice who mixes them together on a 50:50 beamsplitter and then detects both outputs with homodyne detectors, tuned to detect conjugate field quadratures. A classical message containing the outcomes of the detections is sent from Alice to Bob. Bob holds the other half of the entanglement. He uses the classical message to displace each of the quadratures of his half of the entanglement proportionally to the outcomes of Alice's measurements. The sharing of the entanglement allows the quantum state so transmitted to have a similarity to the state that Alice held superior to that which could be achieved by any purely classical channel. Several demonstrations of continuous variable teleportation have been carried out \cite{FUR98,BOW03} with fidelities above 80\% now achieved \cite{MIT08}.

Teleportation is normally discussed (and implemented) with unity gain, of the classical channel, i.e. such that the average values of the quadrature variables of Bob's reconstructed state are equal to those of Alice's original state. However it has been known for some time that useful outcomes can also be achieved using non-unity gain for the classical channel \cite{RAL98, RAL99}. We can describe the initial shared entanglement in the number basis as 
\begin{eqnarray}
|EPR \rangle = \sqrt{1-\chi^2} \; \Sigma_{n=0}^\infty \chi^n |n \rangle |n \rangle
\label{E1}
\end{eqnarray}
where the strength of the entanglement is given by the the parameter $\chi$, with $\chi =0$ corresponding to no entanglement and $\chi = 1$ corresponding to maximal entangement. As described in Ref.\cite{RAL99}. if the gain of the classical channel is chosen to be 
\begin{equation}
\lambda = ({{V-1}\over{V+1}})^2
\label{V1}
\end{equation}
where $V=(1+\chi)/(1-\chi)$ is the anti-squeezing of the entanglement source, then in the Heisenberg picture the output mode of the teleporter under ideal conditions of unit efficiency and pure entanglement is given by
\begin{equation}
\hat a_o = \sqrt{\lambda}\; \hat a + \sqrt{1-\lambda}\;  \hat v
\label{t1}
\end{equation}
where $\hat a$ is the input mode and $\hat v$ is a second mode initially in the vacuum state. The transformation of Eq.\ref{t1} is identical to that induced by passing the mode $\hat a$ through a lossy channel of efficiency $\lambda$. As $V \to \infty$, $\lambda \to 1$ and the output becomes identical to the input.

If the quantum channel between Alice and Bob through which the entanglement is distributed has efficiency $\eta$ then this can be compensated for by further adjusting the gain of the teleporter to $\lambda' = \eta \lambda$ such that now the output mode of the teleporter is given
\begin{equation}
\hat a_o = \sqrt{\eta \lambda}\;  \hat a + \sqrt{1-\eta \lambda}\;  \hat v'
\label{t2}
\end{equation}
where $\hat v' = (\sqrt{\eta(1-\lambda)}\hat v_1 + \sqrt{1-\eta} \hat v_2)/(\sqrt{1-\eta \lambda})$, and $\hat v_2$ is another mode prepared in the vacuum state. The transformation of Eq.\ref{t2} is again identical to that induced by passing the mode $\hat a$ through a lossy channel, this time of efficiency $\eta \lambda$. Now in the limit of maximal entanglement the output becoms equivalent to a lossy channel of transmission $\eta$. Thus teleportation using entanglement distributed in this way through a lossy channel cannot improve the effective efficiency.

{\it Noiseless Linear Amplification}: In order to improve the effective channel between Alice and Bob provided by teleportation we need to distill improved shared entanglement. For EPR entanglement this can be achieved using heralded noiseless linear amplification (NLA). An NLA  implements the number state transformation $|n \rangle \to \sqrt{G} |n \rangle$ \cite{Xia10,Ral09} where $G$ is the amplifier gain ($G>1$). This in turn implies that coherent states are amplified without a noise penalty, i.e. $| \alpha \rangle \to |\sqrt{G} \alpha \rangle$, hence the term noiseless amplification. This transformation is necessarily nondeterministic and the probability of success is state and gain dependent, going to zero for certain state/gain combinations that would otherwise lead to unphysical output states. A generic bound on the probability of success of the NLA is obtained by considering its effect on a Gaussian ensemble of coherent states with variance $V_t$. A successful run of the NLA produces a new ensemble of coherent states, now with variance $V_t' > V_t$. The signal to noise of the conditional state has increased. On average, signal to noise should not increase, thus quite generally $P V_t' +(1-P) < V_t$ or
\begin{equation}
P < {{V_t - 1}\over{V_t' - 1}}
\label{P1}
\end{equation}
where $P$ is the probability of success of the NLA and it has been assumed that the NLA produces vacuum when it fails. 

The NLA can be implemented using linear optics and photon counting \cite{Xia10,Ral09}. A basic amplifying unit is constructed from a simple linear optical network. An ancilla single photon is injected along with the state to be amplified. If a single photon is counted in the ancilla state output the device has succeeded. Provided the output state has an average photon number $\bar n <<1$, then the state will be faithfully amplified. A mutli-path inteferometer can be constructed to fan out the input state over $N$ paths, each of which contains a basic amplifying unit. The state is then reconstructed by recombining the paths. Provided $N>>\bar n$ then a state with $\bar n > 1$ can be amplified. The probability of success of the linear optics construction is
\begin{equation}
P_{lo} = {{\xi}\over{(1 + G)}^N}
\label{P2}
\end{equation}
where $\xi$ is a state dependent normalization factor of order 1. Because $N>>\bar n$, $P_{lo}$ is only comparable to $P$ when the amplified output state has $\bar n << 1$. Amplification via this method is impractical when $\bar n >> 1$ due to the very low probability of success.

Application of the NLA to EPR entanglement can increase the entanglement. We are particularly interested in the case in which the entanglement has experienced loss. According to Ref.\cite{Ral09}, for entanglement distributed via a lossy channel of efficiency $\eta$, if the NLA is applied by Bob to his arm of the entanglement, then a successful run of the noiseless amplifier changes the effective efficiency of the line to
\begin{equation}
\eta_{eff} = {{G \eta}\over{1+(G-1)\eta}}
\label{n1}
\end{equation}
In addition the effective entanglement is increased according to
\begin{equation}
\chi_{eff} = \chi \sqrt{1+(G-1)\eta}
\label{n2}
\end{equation}
Using Eq.\ref{P1} we can conclude that in principle the probability of success for this transformation is bounded by
\begin{equation}
P < {{1- \chi^2(1+(G-1)\eta)}\over{(1+(G-1)\eta)(1-\chi^2)}}
\label{P3}
\end{equation}
\begin{figure}[htb]
\begin{center}
\includegraphics*[width=8cm]{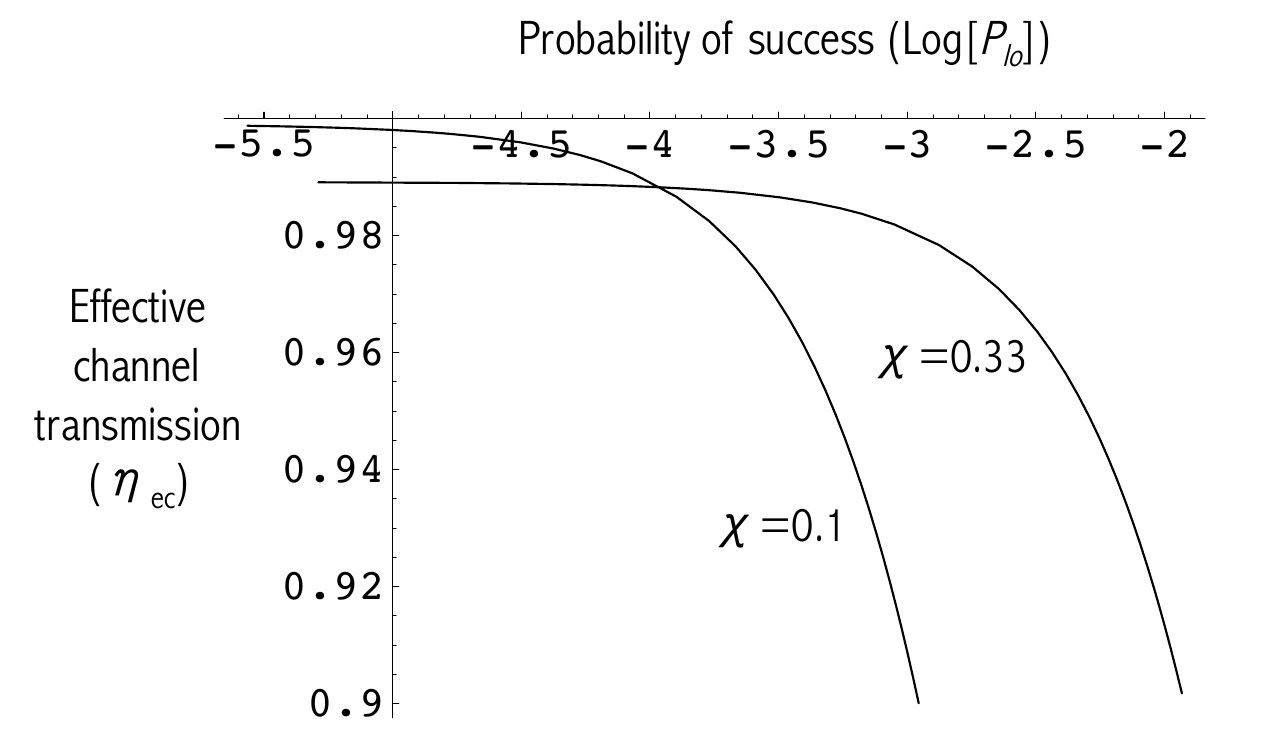}
\caption{Upper bounds on probability of success versus effective channel transmission for the error corrected channel on a Log-linear scale. Regions of the graph below the curves are physically allowed. The initial channel transmission is $\eta =0.9$ but the graphs would be almost identical for lower initial channel transmissions. Two different values for the initial entanglement are plotted.}
\label{fig1}
\end{center}
\end{figure}

{\it Error Correction}: We are now in a position to describe the error correction protocol (see Fig.1(b)). The lossy channel we wish to error correct runs between Alice and Bob. Alice holds a two-mode squeezer, i.e. a source of EPR entanglement, whilst Bob holds a heralded noiseless amplifier. To prepare the channel Alice repeatedly sends entanglement to Bob who tries to amplify it. When Bob succeeds he tells Alice who then uses the distributed entanglement to teleport her quantum state to Bob, with a classical gain $\lambda' = \eta_{eff} \lambda_{eff}$, where $\lambda_{eff}$ is given by Eq.\ref{V1} with the substitution $V \to V_{eff} = (1+\chi_{eff})/(1-\chi_{eff})$. Hence we conclude from Eqs \ref{t2}, \ref{n1} and \ref{n2} that the efficiency of the error corrected channel is
\begin{equation}
\eta_{ec} = G \eta ({{V-1}\over{V+1}})^2 = G \eta \chi^2
\label{n3}
\end{equation}
For $G > 1/\chi^2$ we have $\eta_{ec} > \eta$ and the channel has been error corrected. On the other hand, according to Eq.\ref{P3} $G < (1-(1-\eta)\chi^2)/(\eta \chi^2)$ as the probability of success is zero for higher gains. Thus the best effective channel that can be achieved is
\begin{equation}
\eta_{ecl} =  1-(1- \eta) \chi^2
\label{n4}
\end{equation}
In principle, by making $\chi^2 << 1$, any lossy channel can be made arbitrarily close to an effective channel of ideal transmission ($\eta_{ecl} = 1$). Physically the approach is to distribute very weak EPR entanglement through the lossy channel and try to amplify it up to very strong entanglement with the NLA. When this succeeds a very good effective teleportation channel is created. Of course, Eq.\ref{n4} is in the limit of zero probability of success. In Fig.2 we plot effective transmission against probability of success for two values of $\chi$, showing the trade-off between best effective channel and probability of success.

{\it Error Correction with Linear Optics}: As mentioned earlier, amplification via the known linear optics method is impractical for states of large photon number, or in particular to produce states of large photon number. In order to achieve the effective channel transmissions depicted in Fig.2, amplification to very high squeezing levels, and hence high photon numbers, is required. Hence probabilities of success would be prohibitively low for achieving high transmissions using the linear optics approach. However, Eq. \ref{n3} tells us that even very poor channels can be improved. This may still result in useful error correction even if the final channel remains quite poor. We expect linear optics techniques to be practical in this scenario. This is confirmed by Fig.3 where we plot effective transmission against actual probability of success for the linear optics scheme. Strong initial entanglement allows an order of magnitude improvement in the effective channel transmission (albeit off a low base) with probabilities of success as high as 1\%.
\begin{figure}[htb]
\begin{center}
\includegraphics*[width=8.5cm]{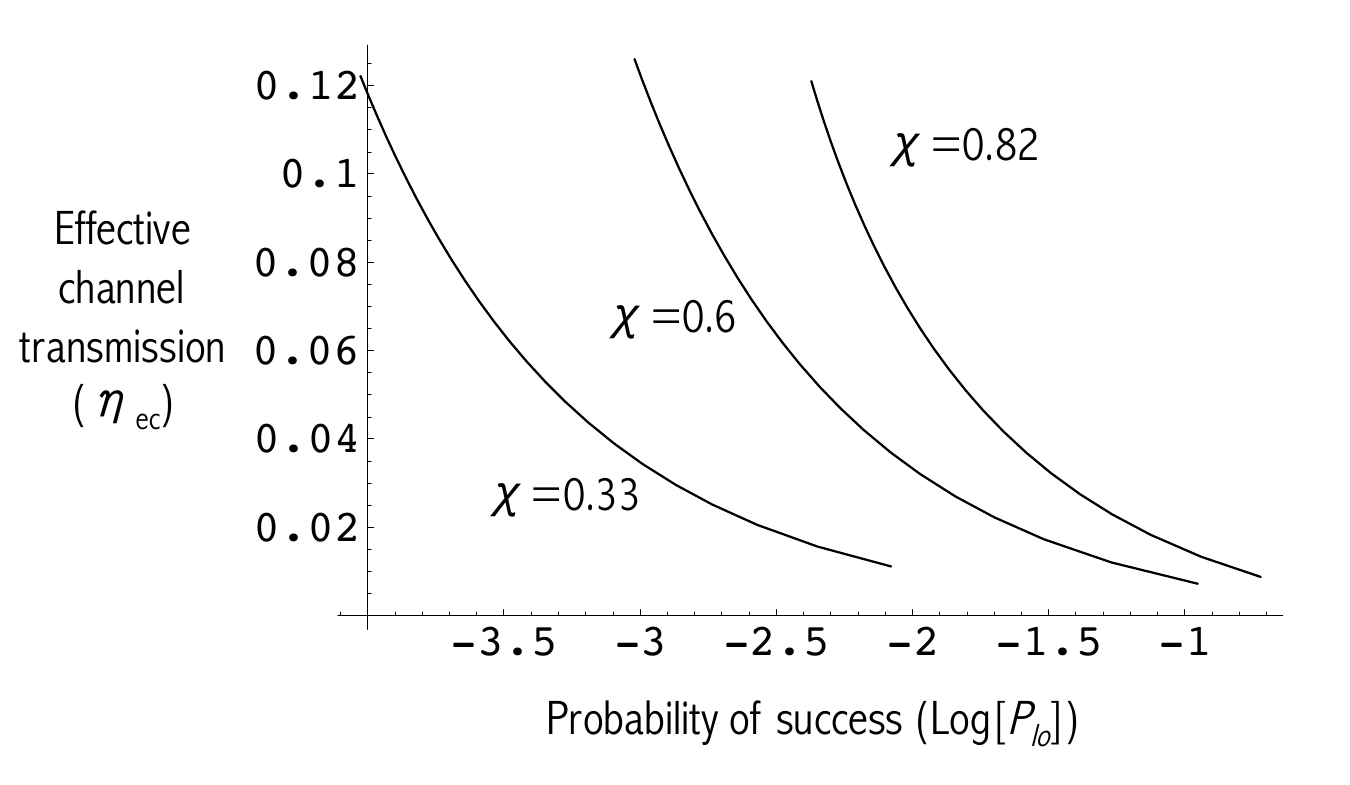}
\caption{Probability of success versus effective channel transmission for the error corrected channel on a Log-linear scale for the linear optics implementation of the NLA. The initial channel transmission is $\eta =0.01$. Three different values for the initial entanglement are plotted corresponding to: 50\% squeezing ($\chi = .33$); 75\% squeezing ($\chi = .60$); and 90\% squeezing ($\chi = .82$). We use $N=2$ which results in fidelities $F>0.995$ between the expected and actual output states for all plots. }
\label{fig1}
\end{center}
\end{figure}

{\it Conclusion}: We have constructed an error correction scheme based on continuous variable entanglement and teleportation, and noiseless linear amplification. The scheme can correct loss induced errors on any field states passing through the channel. This includes qubit states based on single photons and Schr\"odinger cat type states, as well as more traditional continuous variable states such as coherent states. In principle the error correction can be implemented using only linear optics and photon counting and we have shown that significant improvements in effective channel transmission can be achieved using known techniques. Correcting to high channel transmissions is impractical with current techniques due to very low probabilities of success, however the theoretical bounds for the NLA do permit error correction to high channel transmissions for reasonable probabilities of success, motivating a search for more efficient protocols. We have restricted ourselves to pure loss here. A non-trivial extension to this work would be to consider the error correction of thermalized channels.

Throughout this discussion, for simplicity, we have assumed ideal operation of the error correction elements: the entanglement generation; the teleporter; and the NLA. This is a reasonable assumption provided their efficiencies are much higher than that of the channel that is being corrected. Such a scenario is consistent with quantum communications applications. However, for quantum computing applications we require fault tolerance \cite{Sho95, Ste96}: i.e. the elements used to correct the errors may be as inefficient as the channel itself. It is an open question as to whether error correction of the type described here can be made fault tolerant. 

{\it Acknowledgements}: We thank Peter van Loock, Geoff Pryde and Nathan Walk for useful discussions. This research was conducted by the Australian Research Council Centre of Excellence for Quantum Computation and Communication Technology (Project number
CE110001027).


\begin{thebibliography}{99}
\bibitem{note1} Gaussian states have quadrature amplitudes that exhibit a Gaussian probability distribution. Gaussian operations and measurements are linear (or linearizeable) with respect to the quadrature amplitudes. Such operations and measurements maintain the Gaussian character of Gaussian states.
\end{thebibliography}
\end{document}